\documentclass[prb,twocolumn,groupedaddress]{revtex4-1}
\usepackage{amsmath}
\usepackage{bm}
\usepackage{graphicx}
\usepackage{url}
\usepackage{hyperref}
\usepackage{siunitx}
\begin{document}
\bibliographystyle{apsrev4-1}
\title{Trapping of electrons around nanoscale metallic wires embedded in a semiconductor medium}
\author{Chi Cuong Huynh, R.Evrard, and Ngoc Duy Nguyen}
\affiliation{D\'{e}partement de Physique B5, CESAM/Q-MAT, SPIN, Universit\'{e} de Li\`{e}ge, B-4000 Li\`{e}ge, Belgium}
\email{NgocDuy.Nguyen@uliege.be}
%
\begin{abstract}
We predict that conduction electrons in a semiconductor film containing a centered square array of metal nanowires normal to its plane are bound in quantum states around the central wires, if a positive bias voltage is applied between the wires at the square vertices and these latter. We obtain and discuss the eigenenergies and eigenfunctions of two models with different dimensions. The results show that the eigenstates can be grouped into different shells. The energy differences between the shells is typically a few tens of meV, which corresponds to frequencies of emitted or absorbed photons in a range of $\SI{3}{THz}$ to $\SI{20}{THz}$ approximately. These energy differences strongly depend on the bias voltage. We calculate the linear response of individual electrons on the ground level of our models to large-wavelength electromagnetic waves whose electric field is in the plane of the semiconductor film. The computed oscillator strengths are dominated by the transitions to the states in each shell whose wave function has a single radial node line normal to the wave electric field. We include the effect of the image charge induced on the central metal wires and show that it modifies the oscillator strengths so that their sum deviates from the value given by the Thomas-Reiche-Kuhn rule. We report the linear response, or polarizability, versus photon energy, of the studied models and their absorption spectra. These latter show well-defined peaks as expected from the study of the oscillator strengths. We show that the position of these absorption peaks is strongly dependent on the bias voltage so that the frequency of photon absorption or emission in the systems described here is easily tunable. This makes them good candidates for the development of novel infrared devices.
\end{abstract}
%
\maketitle
%
\tableofcontents
%
%
%
\section{Introduction}
\label{int}
The microelectronic industry has achieved great successes since the introduction of integrated circuits. In the past 20 years, the successful advances include critical approaches to boost silicon-based technologies \cite{huang:2001,taraschi:2004dc,Reiche:2007ky} or to introduce silicon-compatible alternatives.\cite{Li:2009ec,Dong:2004,Wang:2011gpa,Wang:2010gg} However, size reduction remains a priority for this industry. At the time of writing this article, several firms are engaged in a race toward the production of integrated circuits with $\SI{7}{nm}$ or less as element dimensions. These dimensions approach the range in which quantum effects become important and connections between elements appear as quantum wires. Therefore, the knowledge of the properties of charge carriers in quantum states inside nanowires has become crucial for the design of future nanoelectronic circuits. Fortunately, one-dimensional (1D) semiconductor nanostructures such as nanowires, nanorods, nanobelts, and nanotubes have for some time been the subject of intense research in both the academic and the industrial worlds so that the quantum properties of charge carriers in these structures are rather well known. See, e.g., Refs.\ \onlinecite{arb-xiong, dayeh, col} and the references therein. An important result concerns the confinement of the charge carriers in quantum states if the wires are thin enough, which allows practical applications in the fields of electronics and optoelectronics.

Curiously, on the contrary, the properties of charge carriers in the outside vicinity of nanowires embedded in semiconductors have attracted little, or even no attention at all. We are aware of only two articles \cite{allan, post} discussing the states of charge carriers in the space of a semiconductor film outside an array of nanostructures, nanoholes in this case. Some other articles describe the state of a 2D electron gas in an array of repulsive potentials constituting a superlattice of obstacles to the electron motion. See, e.g., Ref.\ \onlinecite{allan} and the references therein. The purpose of all these works is to build exotic miniband structures, such as Dirac cones, flat bands, etc. Obviously, with repulsive potentials, there is no bound state in the electron energy spectrum. In the case of nanowires embedded in a semiconductor film, there remain many unanswered questions. Do quantum states form around nanowires? What are their quantum-mechanical properties? How do these properties depend on the wire electric potential and on its size? Do the transitions between these states give rise to resonances in the electromagnetic spectrum? In what range of frequencies? These are important questions whose answers could contribute to the understanding of the quantum mechanics of electrons in semiconductors as well as to the development of new electronic and optoelectronic devices.

The present article is devoted to the study of these questions in the case of charge carriers bound around charged metallic wires inside a semiconductor film. These wires are normal to the film plane and the charge at their surface is due to an applied bias voltage. To ensure the film neutrality, half of these wires are kept at the ground potential. The eigenenergies of the quantum-mechanical states expectably depend on the bias voltage, making the transition frequencies easily tunable. This is in opposition to the case of applications based on transitions inside semiconductor nanowires whose energies are mainly determined by the size and shape of the wires so that changing the transition frequencies by means of an applied voltage is not easily achievable. The relative ease of tuning the transition frequencies in the systems discussed in this article opens a large range of possible optical and electrooptical applications, notwithstanding the difficulty of practical realization. Depending on the wire radius and on the bias voltage, the frequency range extends from the THz domain to the near-infrared one.

Our work is entirely theoretical and mainly exploratory. Its aim is to reveal the interest and importance of the proposed systems. Their practical realization lies beyond our purpose. We focus our attention on models which are described in the next section and whose properties obtained by numerical computation are discussed later on. We first describe the eigenenergy spectrum of single electrons in these models and then devote a detailed discussion to the prediction of their infrared properties, in particular their linear response to an electromagnetic wave and their spectrum of power absorption from this wave. The electron bound in a quantum state induces an image charge in the central metallic wire. We take the effects of this image charge on the interaction with the electromagnetic wave into account and show how it modifies the linear response of the systems under study. It notably changes the sum rule of oscillator strengths, which deviates to some extent from the Thomas-Reiche-Kuhn law.
%
%
\section{Model and methodology}
\label{mod}
%
%
\subsection{Model geometry and characteristics}
\label{geom}
%
\begin{figure}[bt]
\includegraphics{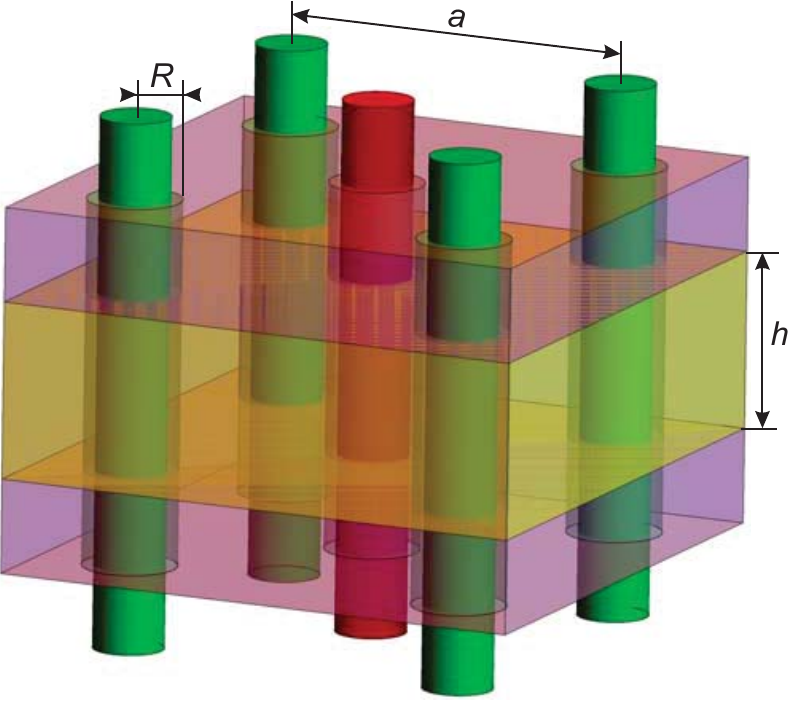}
\renewcommand{\baselinestretch}{1}
\caption{View of a square cell of the wire array. The semiconductor film is represented in orange. It is sandwiched between two insulating layers drawn in purple. The wires in green, located at the square vertices, are at the ground potential. The central wires in red are at the bias potential. $R$: wire radius, including the thickness of the insulating sheath; $a$: square-lattice parameter; $h$: film thickness.}
\label{fig1}
\end{figure}
\normalsize
%

Obviously, the interaction of a single cell with the radiation field is too weak to have noticeable effects in practice. This leads us to consider an array of wires rather than a single isolated one. We choose a regular centered square lattice of parallel metal nanowires embedded in an insulating or lightly doped semiconductor film and aligned perpendicularly to the film surface. They are located at the corners and centers of the square lattice, as shown in Fig.~\ref{fig1}. The semiconductor film is enclosed at the top and bottom by insulating layers in order to keep the charged carriers confined in it.

The wires being made of metal, the electric potential is uniform over their whole surface. Those at the square corners, or vertices, are at a common potential taken as ground potential while those at the square centers, connected together, are at a different potential, which we call the bias voltage. The wires are wrapped in an insulating sheath to avoid charge injection into the film. They are long enough for the electric field to be parallel to the film plane and, therefore, perpendicular to the wires. The wire radius, including the insulating sheath thickness, is denoted by $R$, the film thickness by $h$, while $a$ is used for the superlattice parameter, i.e., the distance between two successive wires along the cell side. We disregard the possible charges on interface states between the insulator sheath and the metal or the semiconductor film. The effect of such charges near the ground wire would be a uniform shift of all the energies, band energies and bound-state energies, without any noticeable result on the properties studied in this article. Interface charges next to the central wires would simply lead to a shift in the bias voltage but would in no way modify the qualitative conclusions of the present work.

In this article, we describe the case of an $n$-type semiconductor with a positive bias voltage applied to the central wires, but, of course, the calculations apply as well to holes in a $p$-type semiconductor, the bias voltage being then negative. The practical realization of such arrays lies beyond the aim of the present article. As proposed in Ref.\ \onlinecite{allan}, cylindrical holes could be drilled in the semiconductor film, either by photolithography, electron-beam lithography, or ion-beam lithography; then the internal surface of the holes would be covered with a thin layer of insulator before finally filling up the holes with a metal or any suitable conductive material.
%
%
\subsection{Formalism and approximations}
\label{form}
Electrons are present in weak concentration in the semiconductor film of the systems under study in this article. They can be due to a light n-type doping. Alternatively, they can be injected either electrically at junctions on the boundaries of the wire array or optically. Their concentration is such that the Fermi level lies relatively close to the ground-state energy of the sates bound to the central wires. In the case of thin wires, so that the eigenfunctions are somewhat compact, the Coulomb repulsive energy between electrons prevents the occupation of the bound states by more than a single electron and we can restrict ourselves to the study of single-electron properties. In the case of thick wires, the average distance between electrons in different eigenstates is large and the Coulomb repulsion is too weak to prevent the occupation of a cell by several electrons. However, the same electrical interaction is also too weak for the correlations between electrons to play an important part in the studied properties, specially in the optical ones. Therefore, the one-electron approximation used in this work seems appropriate to this latter case also. Obviously, the spin-orbit coupling and the relativistic effects are very weak and can be neglected. Throughout this article, we use the well-known effective-mass approximation for the electrons, which is appropriate for the present study. Indeed, the cell dimensions and, therefore, that of the bound-electron wave functions, are large compared to the size of the semiconductor crystal lattice cell. We assume that the conduction band of the semiconductor which the film is made of presents a single minimum at the center of the Brillouin zone. With all these approximations, the sought electron eigenstates $\psi_i(\mathbf r)$ and eigenvalues $\epsilon_i$ are the solutions of the usual Schr\"{o}dinger equation
\begin{equation}
H_0 \psi_i(\mathbf r) = \epsilon_i \psi_i(\mathbf r),
\label{md1}
\end {equation}
in which equation $H_0$ denotes the following Hamiltonian
\begin{equation}
H_0 = - \frac {\hbar^2} {2 m_t} \nabla^2 - e V(\mathbf r).
\label{md1a}
\end {equation}
The notation $m_t$ is used for the electron band-mass tensor, $-e$ for the electron charge, and $V(\mathbf r)$ for the electric-potential distribution in the cell array. The $Ox$ and $Oy$ axes are taken along the square sides in the film plane and the $Oz$ axis normal to this latter. The wires are long enough, or the film surfaces conveniently passivated, so that the electric field within the film is parallel to its plane and the electric potential does not depend on $z$. Most semiconductors used in practice are either cubic or hexagonal so that we restrict ourselves to semiconductor films made of such kinds of material. Moreover, we assume that the semiconductor film has been grown with its $c$ axis normal to the film plane. Therefore, in the Schr\"{o}dinger equation, the directions $x$ and $y$ in the film plane are decoupled from the third direction $z$ and the eigenfunctions can be written as
\begin{equation}
\psi_i(\mathbf r) = \phi_\perp(z) \phi_\parallel(x,y).
\label{md2}
\end {equation}

The part of the wave function depending on $z$, $\phi_\perp(z)$, is that of a free band electron moving along a crystal symmetry axis and confined in a narrow space between two plane boundaries. The study of this part of the wave function is a very well known problem. It is discussed in several textbook on the quantum mechanics of nanostructures. Its solutions are also those for the confinement of electrons in quantum wells. Chapter 4 of Ref.\ \onlinecite{nag}, e.g., gives more information on these questions. The solution depends on the confining potential at the film surfaces. In the case of steep and totally reflecting interfaces, it is written as
\begin{equation}
\phi_\perp(z) = \sqrt{\frac {2} {h}} \sin \kappa_{n_z} z
\label{md3}
\end {equation}
with $\kappa_{n_z} = n_z \pi / h$, $n_z = 1,2,3, \cdots$ being the quantum number for the $z$ direction. The corresponding $z$ contributions to the eigenenergies are $\epsilon_{n_z} = \hbar^2 \kappa_{n_z}^2/ 2 m_z$, $m_z$ denoting the $c$ component of the diagonal band-mass tensor. The study of this part of the wave function has no interest for our work. Therefore, from now on, we disregard it completely as well as its contribution to the eigenenergies and we are left with finding the eigenfunctions and eigenvalues of the 2D Schr\"{o}dinger equation
\begin{equation}
- \frac {\hbar^2} {2 m_b} \nabla^2 \phi_{\parallel,i}(\mathbf r) - e V(\mathbf r) \phi_{\parallel,i}(\mathbf r) = \epsilon_i \phi_{\parallel,i}(\mathbf r)
\label{md4}
\end {equation}
in which $m_b$ is now the electron band mass for its motion perpendicular to the semiconductor $c$ axis and $\mathbf r$ a two-component vector, $\mathbf r = (x,y)$.

Obviously, due to the translational invariance, the eigenfunctions are 2D Bloch functions, i.e.,
\begin{equation}
\phi_{\parallel,i}(\mathbf r) = u_{i,\mathbf k} (\mathbf r) e^{i \mathbf k \cdot \mathbf r}.
\label{md5}
\end {equation}
In this equation, Eq.\ (\ref{md5}), $\mathbf k$ is a 2D reciprocal vector of the superlattice under study; the factor $u_{i,\mathbf k} (\mathbf r)$ is periodic with the same periodicity as the wire array and obeys the equation
\begin{eqnarray}
- \frac {\hbar^2} {2 m_b} \nabla^2 u_{i,\mathbf k} (\mathbf r) -i \frac {\hbar^2} {m_b} \mathbf k \cdot \bm{\nabla} u_{i,\mathbf k} (\mathbf r) - e V(\mathbf r) u_{i,\mathbf k} (\mathbf r)
\nonumber
\\
= \left(\epsilon_{i,\mathbf k} - \frac {\hbar^2 k^2} {2 m_b} \right) u_{i,\mathbf k} (\mathbf r).
\label{md6}
\end{eqnarray}
It is well known that the energy eigenvalues in superlattices form energy bands, often called minibands, functions of $\mathbf k$. Energy bands in 2D lattices have been the subject of many studies. See Ref.\ \onlinecite{allan} and the references therein for more information. In particular, this latter article discusses the minibands in the case of films with a regular array of cylindrical holes bored in it. Those systems have some similarities with what we propose in the present article. However, in their case, as the holes are empty, there is no attractive electric potential acting on the electrons and no bound states are formed. Indeed, that work is devoted to the study of the energy minibands of electrons freely moving in the film between the holes and, in particular, to the possible existence of exotic band features like Dirac cones, flat bands, etc. On the contrary, in the present article, we are interested in the state of electrons bound around charged metal wires filling the holes. We leave the study of the energy minibands of our wire arrays for a future work and restrict ourselves to the case $\mathbf k = 0$. In fact, in the examples we treat in detail, the overlap between the ground-state wave functions of two neighboring cells is weak and the electrons are localized around the central wires. For the sake of conciseness, we drop the index 0 in the periodic part of the wave function and simply write $u_i (\mathbf r)$. As a conclusion, the sought eigenfunctions and eigenstates are the solutions of the simple 2D partial differential equation.
\begin{equation}
- \frac {\hbar^2} {2 m_b} \nabla^2 u_i (\mathbf r) - e V(\mathbf r) u_i (\mathbf r) = \epsilon_i u_i (\mathbf r).
\label{md7}
\end {equation}
%

%
\subsection{Symmetry considerations}
\label{sym}
Of course, the solutions of Eq.\ (\ref{md7}) must be compatible with the symmetry of the square. In Cartesian coordinates, it is quite natural to choose solutions of Eq.\ (\ref{md7}) with a symmetry appropriate to reflections onto the $Ox$ and $Oy$ axes. As these symmetry operations commute, we can classify the solutions in four groups with wave functions totally symmetric, totally antisymmetric, and symmetric with respect to an axis while antisymmetric to the other, respectively. Obviously, the solutions of the last two groups are degenerate two by two.

Very near the central wire, the potential is almost cylindrical. This leads to introduce two quantum numbers $n$ and $m$, besides precising the reflection symmetries, to characterize the eigenstates of electrons strongly bound to the wires. These quantum numbers correspond to the principal and magnetic quantum numbers used in problems with a cylindrical symmetry. This is discussed in more detail in Sec.~\ref{eigfen}.
%
%
\section{Eigenfunctions and eigenenergies}
\label{eigfen}
%
%
%
\subsection{Calculation details and results}
\label{res}
Finding the eigenfunctions and eigenenergies of Eq.\ (\ref{md7}) first requires the knowledge of the 2D electric-potential distribution $V(\mathbf r)$. We obtain it by solving the 2D Laplace equation inside a cell with periodic boundary conditions on the square sides opposite to each other and Dirichlet conditions on the wires bringing them either to the ground potential or to the bias potential, depending on the wire position. The Laplace equation is solved using the AC/DC module of the COMSOL Multiphysics platform \cite{coms}. In this computation we use the value of the dielectric constant of GaN in the dielectric film and that of HfO\textsubscript{2} for the insulating sheath around the metal wires but, of course, these values are not really essential for our calculations, which can apply to other semiconductors and insulators as well.

This solution of the Laplace equation is introduced into the partial differential equation, Eq.\ (\ref{md7}), which is then solved by means of the Mathematics module of the same COMSOL platform. Dirichlet conditions are applied to all the wires, imposing that the wave function vanish on their boundary. For the non-degenerate solutions, we again impose periodic conditions on the cell side. However, in the case of degenerate eigenfunctions, the solutions found by COMSOL are not those with a definite symmetry with respect to reflections onto the $Ox$ and $Oy$ axes, but linear combinations of these latter. Then, it is better to integrate Eq.\ (\ref{md7}) on a quarter of the cell with either Dirichlet or Neumann boundary conditions on the axes, depending on the sought symmetry for the solution. For $m_b$, we use the value of the band mass of the GaN conduction band in the plane normal to the $c$ axis, $m_b = 0.2 \, m_0$, $m_0$ being the electron mass in vacuum.

\begin{figure}[t]
\includegraphics{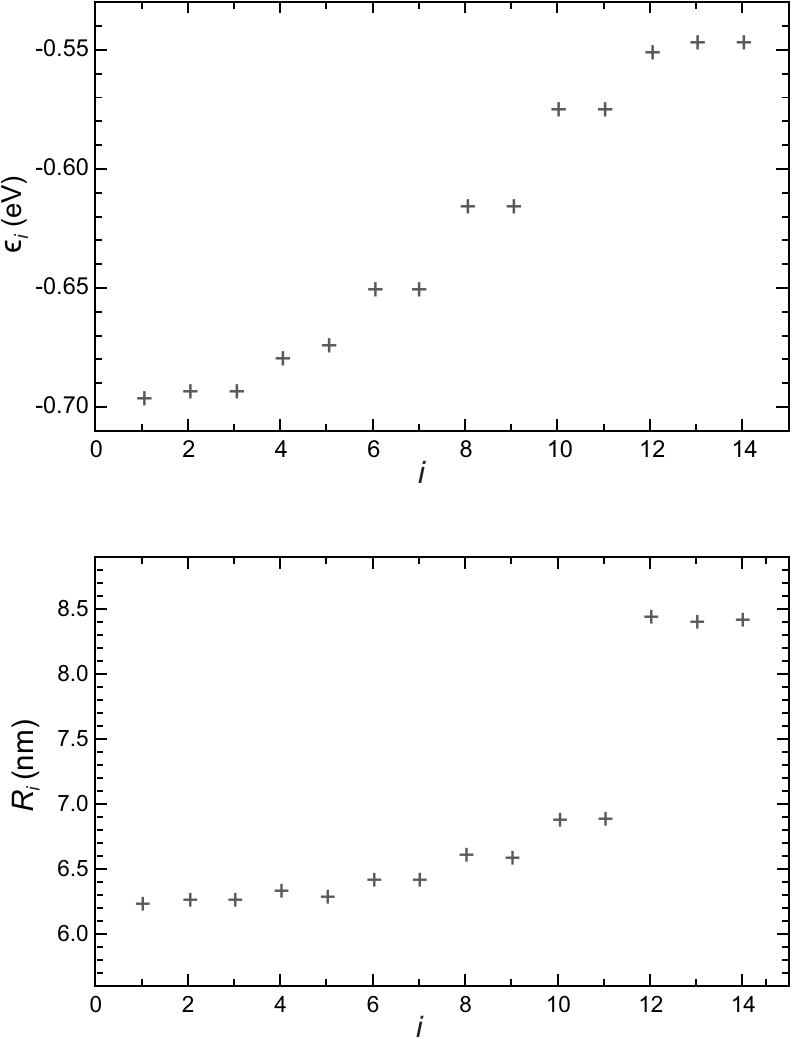}
\renewcommand{\baselinestretch}{1}
\caption{Top: First 14 eigenenergies $\epsilon_i$ of the array with $R = \SI{4}{nm}$ and $a = \SI{30}{nm}$ vs the solution number, $i$. The bias voltage is $V_B = \SI{1}{V}$. Notice the parabolic dependence of the energies on $i$ and the degeneracy of numerous eigenstates. Bottom: Average radii $R_i$ of the electron charge distribution in the same eigenstates, also vs $i$. Two groups of eigenstates are revealed with notably different radii, $R_i \approx \SI{6.5}{nm}$ and $R_i \approx \SI{8.5}{nm}$, respectively. These groups correspond to the electron shells in atomic physics and are characterized by different numbers of wave-function radial nodes and antinodes.}
\label{eigvR}
\end{figure}
\normalsize
%

%
\begin{figure}[bt]
\includegraphics{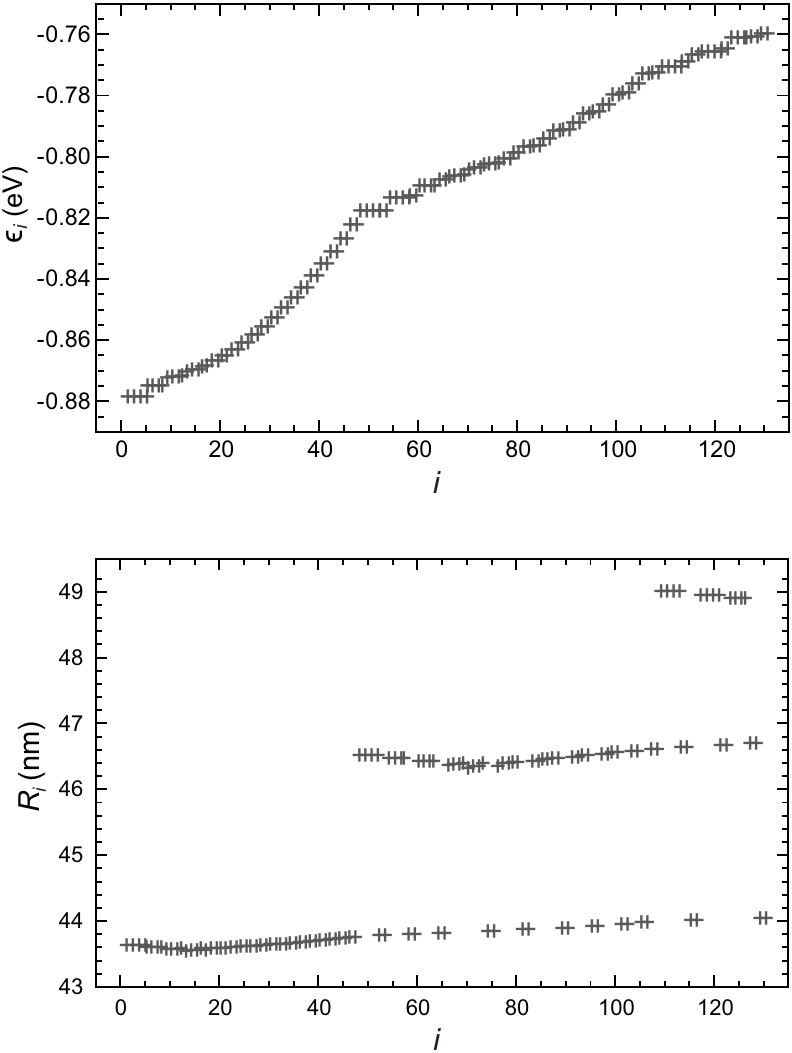}
\renewcommand{\baselinestretch}{1}
\caption{Top: First 130 eigenenergies $\epsilon_i$ of the array with $R = \SI{40}{nm}$ and $a = \SI{200}{nm}$ vs the solution number, $i$. The bias potential is $V_B = \SI{1}{V}$. The states are tightly packed in energy, forming a near continuous distribution. Bottom: Average radii $R_i$ of the electron charge distribution in the same eigenstates, also vs $i$. Three separate shells are visible, with $R_i \approx \SI{43.5}{nm}$, $\SI{46.5}{nm}$, and $\SI{49}{nm}$, respectively. Notice the parabolic energy distribution in the different shells.}
\label{eigvRlg}
\end{figure}
\normalsize
%
In this article, we discuss two models with quite different dimensions. In the first one, the cells are small with a square-lattice parameter $a = \SI{30}{nm}$ and a wire radius $R = \SI{4}{nm}$. The thickness of the insulating layer between the metal wires and the semiconductor film is $\SI{1}{nm}$. In the second model, the cells are larger, $a = \SI{200}{nm}$, and $R = \SI{40}{nm}$. The insulating-layer thickness is now $\SI{5}{nm}$. In all cases, the applied voltage bias, i.e., the voltage difference between the central wire and those at the vertices, is $V_B = \SI{1}{\volt}$. Figure~\ref{eigvR} gives the first 14 eigenenergies and average electron-charge radii of model 1 versus $i$, the eigenstate number in order of increasing eigenenergies, while Fig.~\ref{eigvRlg} gives the same characteristics for the first 130 eigenstates of model 2. The average charge radius is defined as $R_i = \iint (x^2 + y^2)^{1/2} u_i^2(x,y) dx \, dy$, the integral being taken over the surface of a square cell and the square of the norm of $u_i(x,y)$ being normalized to 1 in the same area. Recall that this concerns the part of the wave function depending on $x$ and $y$ only. As stated above, the dependence on $z$ is not discussed in the present article.

The examination of the graphs giving the electron-charge radii shows that the eigenstates can be grouped into different shells as in atomic physics. The difference in radius between the first two shells is clearly visible in Fig.~\ref{eigvR} for model 1, while three shells can be seen in Fig.~\ref{eigvRlg} for model 2. The difference in energy between eigenstates in different shells depends on the wire radius and the bias voltage but is typically a few tens of meV. Inside a shell, the radii have almost the same value for all the wave functions and the energy distribution is almost continuous, especially in the case of model 2. The larger the wire radius, the closer to forming continuous distributions are the eigenenergies belonging to the same shell. In the case of model 2, $R = \SI{40}{nm}$, the energy difference between two successive levels is just a few meV. In this range of energy differences, any perturbation, such as the electron-electron interaction, the interaction with phonons or with crystal defects and impurities, leads to mixing the electron states which are close in energy. Then, the electrons actually occupy linear combinations of the eigenstates obtained here. However, the solutions of the unperturbed Schr\"{o}dinger equation constitute a complete basis of wave functions that can be used in the calculations of the wire-array properties, in particular of the optical ones which are an important aspect of our study. This justifies our interest in the eigenstate spectrum, even in the case of large radii as in model 2.

%
\begin{figure}[bt]
\includegraphics{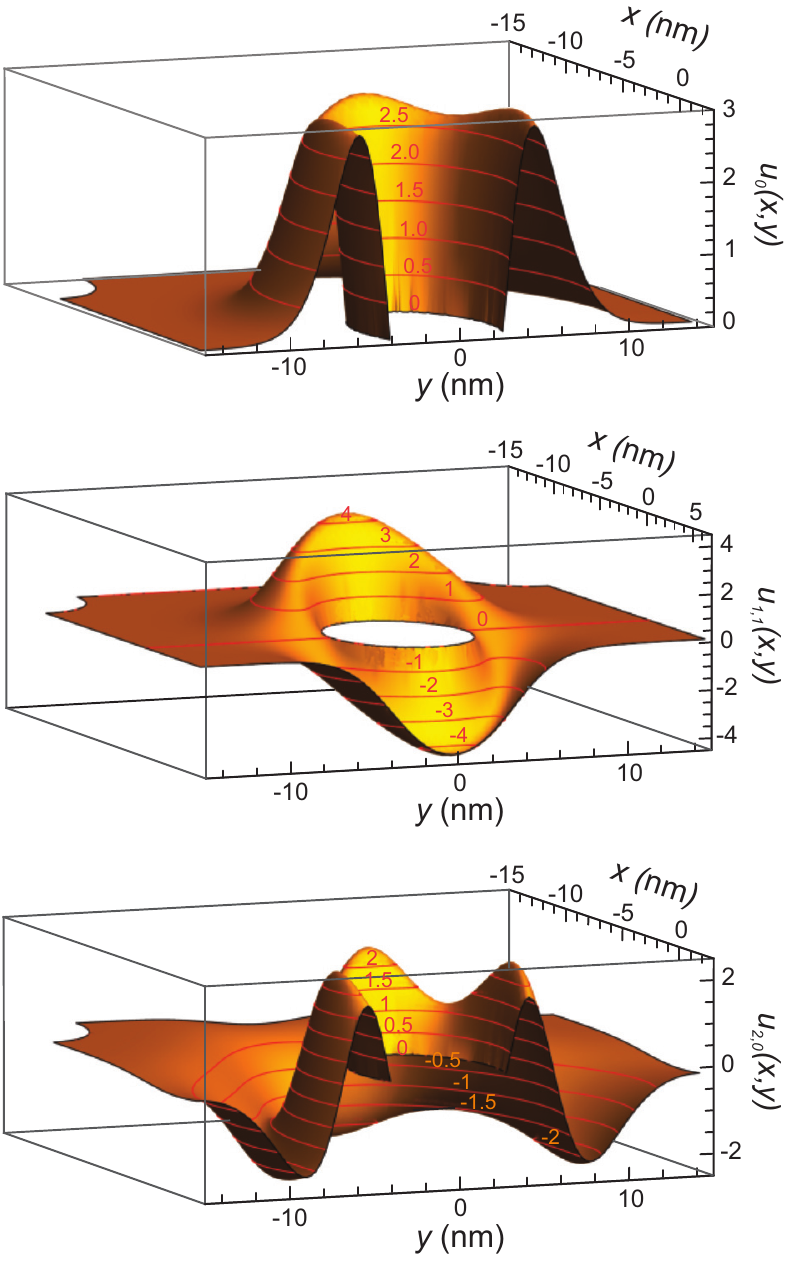}
\renewcommand{\baselinestretch}{1}
\caption{Cut views of 3 wave functions in the case of array model 1 ($R = \SI{4}{nm}$ and $a = \SI{30}{nm}$). Top: ground state, no node; middle: first excited state ($m = 1$) in the first shell of states ($n = 1)$, a single radial node along the $Oy$ axis; bottom: first state in the second shell of states ($n = 2, m = 0$), two radial antinodes. In red, curves of wave-function equal values. See text for more detail, including definition of wave-function normalization.}
\label{wfthinw}
\end{figure}
\normalsize
%
%
\begin{figure}[bt]
\includegraphics{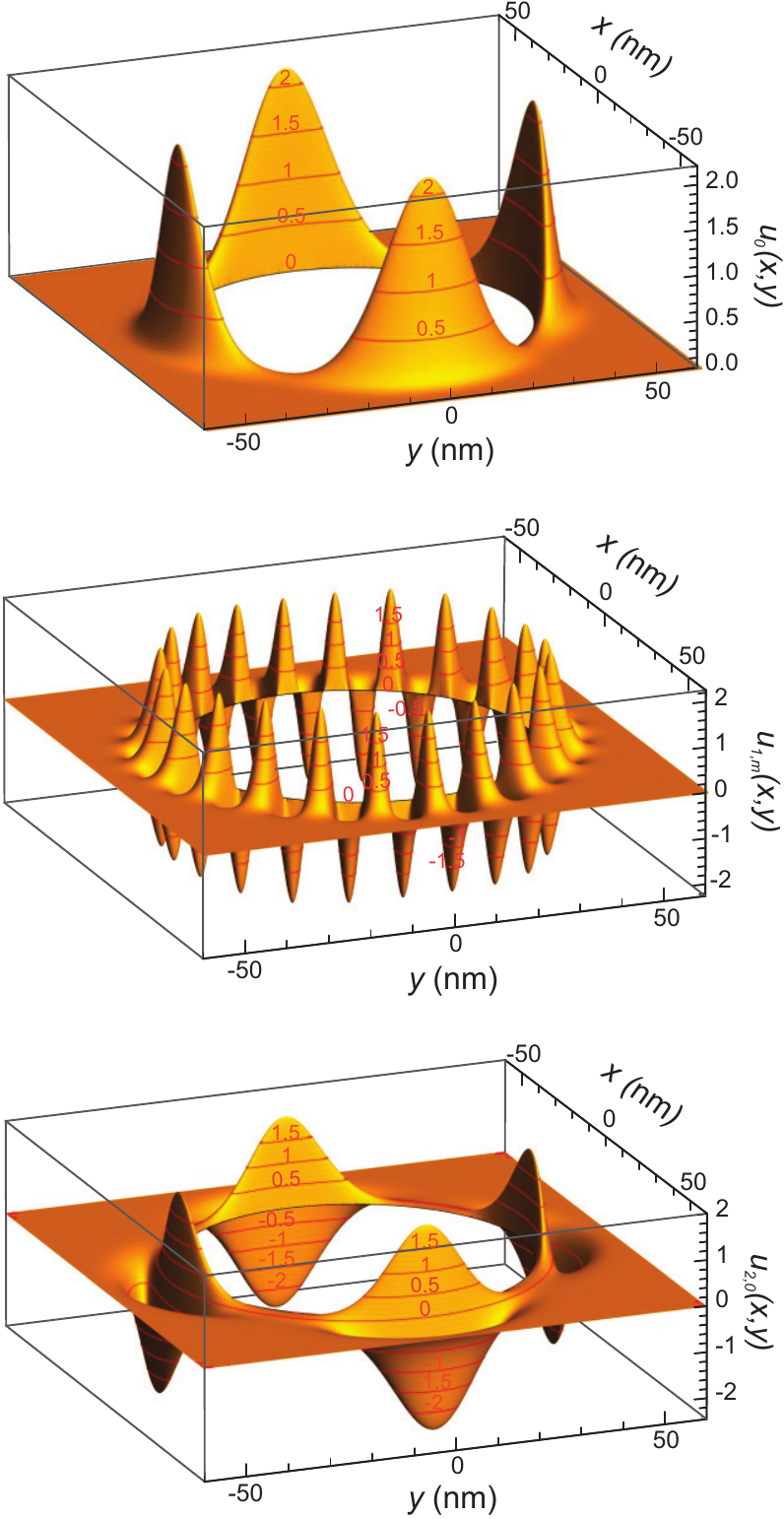}
\renewcommand{\baselinestretch}{1}
\caption{Views of three wave functions in the case of array model 2 ($R = \SI{40}{nm}$ and $a = \SI{200}{nm}$). Top: ground state, no node; middle: highly excited state of the first shell of states, 22 circumferential nodes; bottom: first state in the second shell of states ($n = 2$), a radial node. In red, curves of wave-function equal values, indicated on the curves. See text for more detail, including definition of wave-function normalization.}
\label{wflgw}
\end{figure}
\normalsize
%

The reasons for the eigenfunction properties discussed here become clear after examination of the next figures. Figure~\ref{wfthinw} shows the wave functions in the ground state, the first excited state of the first shell, and the first state of the second shell in the case of model 1. To make the details of the wave functions more visible, the figures are cut by a vertical plane $x = x_0$, with $x_0 = \SI{2}{nm}$ for the first and last graphs and $x_0 = \SI{6}{nm}$ for the middle one, only the part with $x < x_0$ being represented. Figure~\ref{wflgw} shows the wave functions in the ground state, a highly excited state of the first shell, and the first state of the second shell in the case of model 2, in a part of the cell relatively close to the central wire. In both models, the bias voltage is $\SI{1}{V}$. In the figures discussed here, the lines in red are the curves of equal values of the eigenfunctions, values indicated in red on the curves. The eigenfunction normalization used in drawing Figs.~\ref{wfthinw} and \ref{wflgw} is that of COMSOL \cite{coms} and, therefore, is adapted to the computation on a discrete mesh of points by means of the finite-element method. It appears as arbitrary for our purpose and most of the calculations in the next sections require that the square of the eigenfunctions be renormalized to 1 on the cell area.

As demonstrated by the present figures, wave functions in different shells have a different number of radial nodes and antinodes while those in a same shell differ by the number of circumferential nodes and antinodes. This leads us to label the eigenfunctions with two indices; the first one, $n = 1, 2, \ldots$, gives the number of radial antinodes specific to the shell to which the eigenfunction belongs and, therefore, plays the role of the principal quantum number in atomic physics. The second one, $m = 0, 1, 2, \ldots$, denotes the number of the wavefunction circumferential nodes. For the sake of conciseness, we label the ground state with the single index 0. At low bias voltage, the electrons are no longer strongly bound to the wires and the wave-functions in adjacent cells have a large overlap. In this case, the notion of circumferential nodes and antinodes becomes somewhat blurred. Notice that an isolated center has a cylindrical symmetry; the quantum numbers used here are obviously appropriate to this case also; then $m$ becomes the electron magnetic quantum number, the negative values of which are in relation with the way the degeneracy of states is taken care of.

Due to the square symmetry of the cell, the eigenfunctions can be built either symmetric or antisymmetric with respect to reflections onto two directions orthogonal to each other, e.g., that of the $Ox$ and $Oy$ axes. Making the distinction between states in the same shell and having the same number of circumferential nodes requires to specify the symmetry with respect to an axis, at least. For example, the eigenfunction represented in the middle graph of Fig.~\ref{wfthinw} is symmetric with respect to reflections onto the $Ox$ axis and antisymmetric with respect to those onto the $Oy$ axis. It is degenerate with the eigenfunction obtained from it by a $\SI{90}{\deg}$ rotation, which has the opposite symmetry.

%
\begin{figure}[t]
\includegraphics{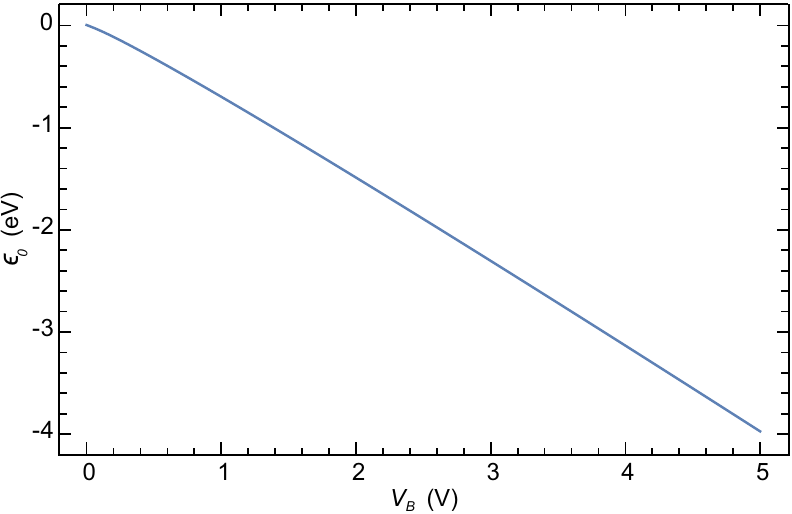}
\renewcommand{\baselinestretch}{1}
\caption{Ground-state energy of model 1 vs bias voltage. Their relation is almost linear with a slope of about $-4/5$.}
\label{eps0V}
\end{figure}
\normalsize
%
The existence of circumferential nodes and antinodes reflects the formation of stationary waves by electrons rotating around the wire. This allows the following back-to-the-envelope calculation which sheds some light on the eigenenergy dispersion relation. Call $R_n$ the average radius of the wave functions in the $n$ shell. The wavelength of a circular stationary wave around the wire with $m$ circumferential nodes is
\begin{equation}
\lambda_m = \frac {2 \pi} {m} R_n
\label{r1}
\end {equation}
so that its kinetic energy is
\begin{equation}
E_K(m) = m^2 \frac {\hbar^2} {2 m_b R_n^2}.
\label{r2}
\end {equation}
As the states in a shell are all at about the same depth in the electric potential, they have nearly the same potential energy. Therefore, the dependence of the energy eigenvalues $\epsilon_{n,m}$ on $m$ is governed by the expression of the kinetic energy of Eq.\ (\ref{r2}). This explains the parabolic aspect of the energy curves in the upper parts of Figs.~\ref{wfthinw} and ~\ref{wflgw}.
%
\begin{figure}[t]
\includegraphics{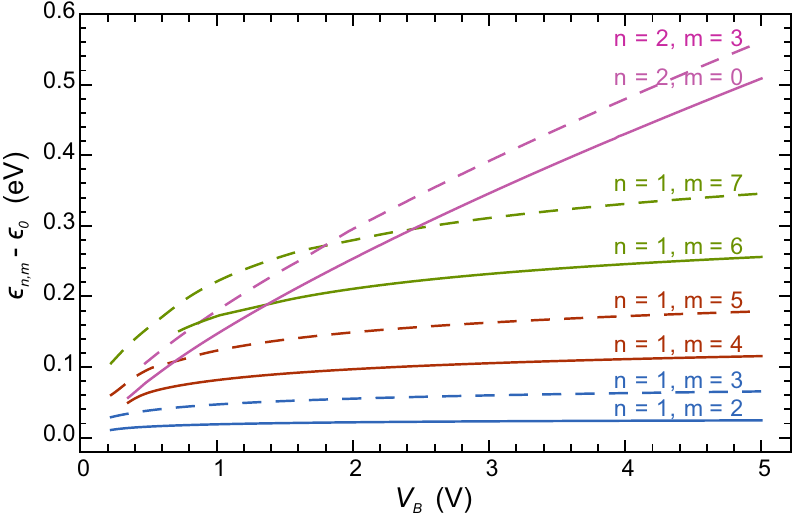}
\renewcommand{\baselinestretch}{1}
\caption{Excitation energies of several low-order eigenstates of model 1 vs bias voltage. Full lines: Eigenfunctions symmetric with respect to $Ox$ and $Oy$. Dashed lines: eigenfunctions symmetric to $Ox$ and antisymmetric with respect to $Oy$. The eigenenergies of the $n = 2$ states depend more strongly on the bias voltage than those with $n = 1$. See text for the meaning of the quantum numbers $n$ and $m$.}
\label{excV}
\end{figure}
\normalsize
%
%
\subsection{Effect of bias voltage}
\label{biasef}
The possible applications of the systems studied in this article as tunable emitters or detectors of radiation in the infrared range, from THz frequencies to the near infrared, explain our interest in the dependence of the transition energies on the bias voltage. Consider the case of model 1 with wire radii of $\SI{4}{nm}$ and square-lattice parameters of $\SI{40}{nm}$. Figure~\ref{eps0V} shows that the ground-state energy versus the bias voltage decreases almost linearly with increasing bias, the slope being about $-4/5$. Figure~\ref{excV} gives the excitation energies, $\epsilon_{n,m} - \epsilon_0$, of a few eigenstates in the first ($n = 1$) and second ($n = 2$) shells versus the bias voltage in the range between a few tenths of volt and $\SI{5}{V}$. The full lines represent the results for totally symmetrical eigenstates while the dashed ones give those for eigenstates symmetric with respect to reflections onto an axis, for example $Ox$, and antisymmetric with respect to the other one, $Oy$. The curves for states in the same shell become more or less parallel at bias voltage high enough. Therefore, the shift in energy due to changing the bias voltage is then almost the same for all the states in the same shell. The eigenenergies of states in high-order shells, i.e, with values of the quantum number $n > 1$, depend more strongly on the bias voltage. This is an interesting property as the possible optical applications would probably involve transitions between states with different values of $n$.

\begin{figure}[t]
\includegraphics{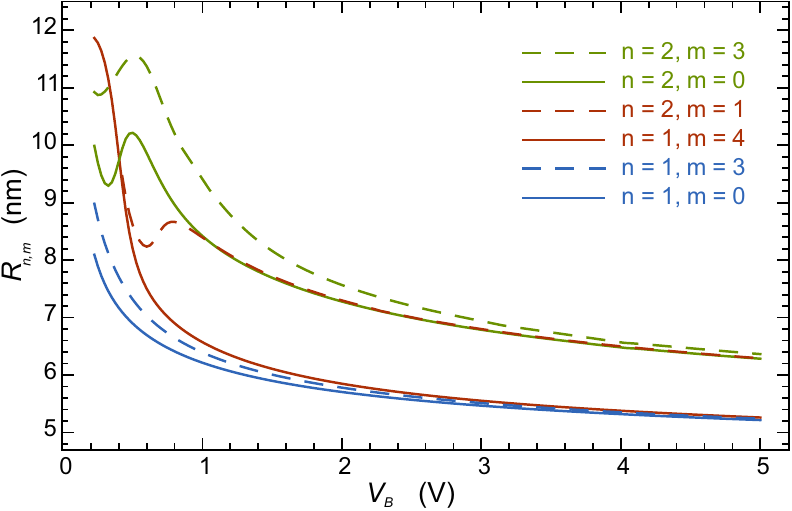}
\renewcommand{\baselinestretch}{1}
\caption{Average wave-function radius of several low-order eigenstates of model 1 vs bias voltage. Full lines: Eigenfunctions symmetric with respect to $Ox$ and $Oy$. Dashed lines: eigenfunctions symmetric to $Ox$ and antisymmetric with respect to $Oy$.}
\label{rvsvb}
\end{figure}
\normalsize
%
%
It is obvious that the average distance of the electron from the central cylinder axis, i.e., the average eigenfunction radius, must decrease with increasing bias. This is confirmed by Fig.~\ref{rvsvb}, which shows this average radius for a few low-order states versus the bias voltage. The behavior of the eigenstates in the shells with $n > 1$ deviates from this rule at low bias voltages. In this case, the wave function is not localized next to the central wire but extends up to the boundary between the adjacent cells. Then, a limited increase in the bias voltage redistributes the wave function along the cell boundaries rather than bringing it closer to the central wire. As apparent from Fig.~\ref{rvsvb}, the behavior of the average radius versus the bias voltage is similar for the eigenfunctions in the same shell. Obviously, this is in direct relation with the similar behavior of the excitation energy revealed in Fig.~\ref{excV}.

The maximum electric field in the structure described here is $\SI{1.094}{MV/cm}$ per volt of applied bias. This is to be compared to the breakdown field of GaN, which is on the order of $\SI{5}{MV/cm}$. This makes GaN a good candidate for building the structures studied in the present article, although the possible existence of a spontaneous polarization in this semiconductor could pose a problem. We have not considered this point in the present work. The breakdown field in other semiconductors is in general far lower. For example, in the case of GaAs, AlAs, and the mixed crystals made of them, the value of the breakdown field lies between 0.4 and $\SI{0.6}{MV/cm}$. Making use of these semiconductors would restrict the bias voltage to values far lower than with GaN or would require systems of larger dimensions.

\begin{figure}[t]
\includegraphics{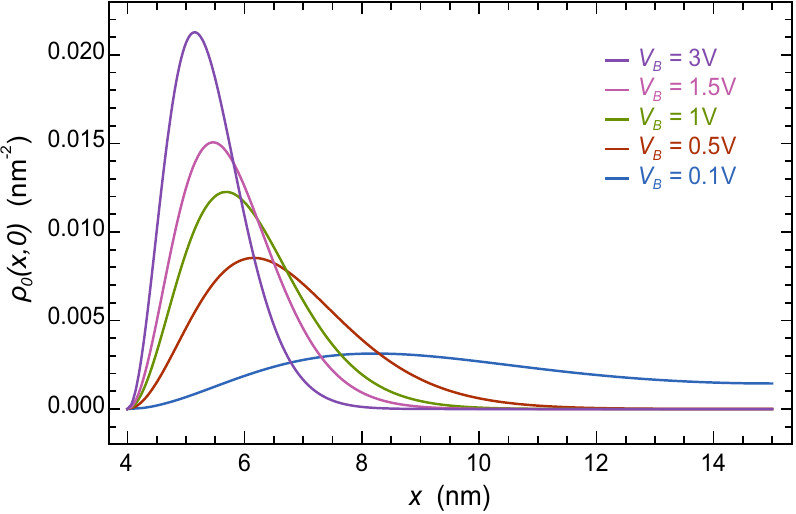}
\renewcommand{\baselinestretch}{1}
\caption{Ground-state two-dimensional electron density along the positive $Ox$ axis for model 1 at bias voltages of $\SI{0.1}{V}$, $\SI{0.5}{V}$, $\SI{1}{V}$, $\SI{1.5}{V}$, and $\SI{3}{V}$.}
\label{edens}
\end{figure}

Of course, the behavior of the electron charge distribution as a function of the bias voltage follows that of the wave function and the electron charge density concentrates closer to the central wire with increasing bias voltage. This is apparent in Fig.~\ref{edens} which shows the two-dimensional electron density of the ground state along the positive part of the $Ox$ axis, $\rho_0(x,0)$, for five bias voltages from $\SI{0.1}{V}$ to $\SI{3}{V}$. The electron surface concentration is in \si{nm^{-2}}. The lowest curve shows that the electron charge distribution in the ground state does not vanish at the cell boundary for a bias of $V_B = \SI{0.1}{eV}$. For shells with $n > 1$, this is true even for larger bias voltages.

%
\normalsize
%
%
\begin{figure}[b]
\includegraphics{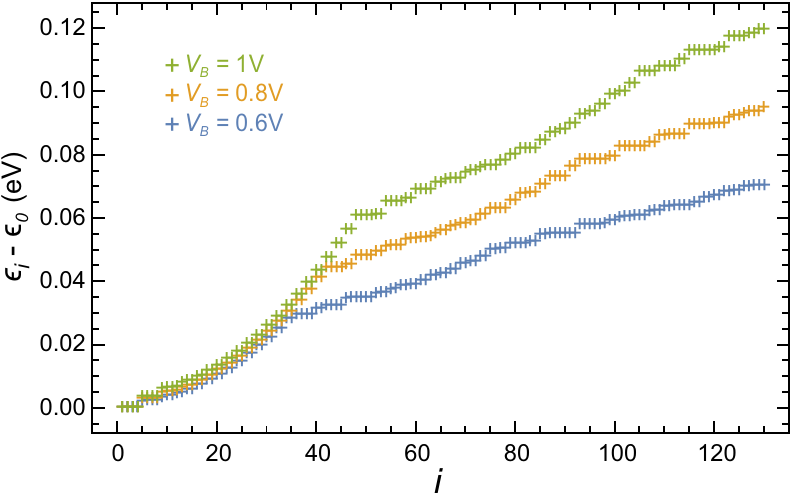}
\renewcommand{\baselinestretch}{1}
\caption{Excitation energies, $\epsilon_i - \epsilon_0$, in the case of model 2 ($R = \SI{40}{nm}$ and $a = \SI{200}{nm}$) vs the solution order, $i$, for three bias voltages. Lower curve: $V_B = \SI{0.6}{V}$; middle curve: $V_B = \SI{0.8}{V}$; upper curve: $V_B = \SI{1}{V}$.}
\label{eigV3b}
\end{figure}
\normalsize
In the case of larger structures, at comparable values of bias voltage, the electrons in general lie farther from the cell boundaries. As a consequence, the property that a change in the bias voltage has almost the same effect on the energies of all the eigenstates in the same shell remains true and is even reinforced in the large structures. This is illustrated in Fig.~\ref{eigV3b}, which shows the excitation spectrum of the second model ($R = \SI{40}{nm}$ and $a = \SI{200}{nm}$) as a function of the index $i$ of the eigenstates ordered in increasing energies, for three bias voltages $\SI{0.6}{V}$, $\SI{0.8}{V}$, and $\SI{1}{V}$. The ground-state energy, not represented in the figure, strongly depends on the bias, as expected. Its values are $\epsilon_0 = \SI{- 304.1}{meV}$, $\SI{-553.9}{meV}$, and $\SI{-879.3}{meV}$ for the bias voltages used in the graphs of Fig.~\ref{eigV3b}, respectively. This figure shows that the curves corresponding to different bias voltages either nearly coincide for the first shell, or are parallel for the next ones. This confirms that the shift in energy produced by a bias-voltage change has about the same value for all the states in a given shell.

%
\section{Linear response of an isolated center to a quasi-static oscillating electric field}
\label{linear}
%
%
\subsection{Motivation and description of the problem}
\label{mot}
The systems proposed in this article being new, their electrical and optical properties are totally unknown. They could however be of great practical interest. Indeed, the energy of excitation between two different shells of levels is in the range of $\SI{10}{meV}$, or a few tens of meV at most. This corresponds to photon frequencies in the THz range, where emitters and detectors are not numerous. A possible application as radiation detector could use the modification of the in-plane electrical conductivity brought by an electromagnetic wave. Indeed, the overlap between wave functions on neighboring sites depends on the electron state of excitation which is changed by the incident wave. Another possible process is the ionization of the bound states which could induce a capacitive current in the central wire.

The following sections describe the linear response of the bound electrons to an ac electric field in the case of the two models introduced previously in Sec.~\ref{eigfen}. This electric field is either a uniform field applied to the whole structure or that of a large-wavelength electromagnetic wave. In model 1, with wire radii $R = \SI{4}{nm}$ and a dc bias voltage of $\SI{1}{V}$, the electrons are rather strongly bound to the central wire in all directions and the response is due to quantum-mechanical transitions, either real or virtual. In model 2, with wire radii of $R = \SI{40}{nm}$ and in the same range of bias voltage, the electrons are almost free to move tangentially around the wire but not in the radial direction. Then, the electron states are similar to those in a quantum well except that the symmetry is almost cylindrical rather than planar. At relatively low frequencies, a mobility $\mu_t$ could be associated with the electron tangential motion and computed, e.g., in the framework of a Drude model. This model relates mobility and collision time $\tau_c$ through the relation $\tau_c = m_b \mu_t / e$. Recall that $m_b$ denotes the electron band mass. Assuming that the mobility is on the same order of magnitude as in the semiconductor bulk and using the values given in Ref.\ \onlinecite{nsm} for samples of high quality, we obtain $\tau_c = \SI{0.3}{ps}$ for GaAs and $\tau_c = \SI{0.05}{ps}$ for GaN at room temperature. At $\SI{100}{K}$, the collision time is at least five times larger. The study of the \textit{\`a la Drude} motion of the electrons lies outside the scope of the present work and we focus on the quantum-mechanical response of the systems under consideration, which is expected to be dominant at frequencies larger than $1/\tau_c$, i.e., a few THz in our case. The two models discussed in this article, with a bias voltage of $\SI{1}{V}$, show inter-shell transitions at frequencies fulfilling this condition.

In the THz range, the radiation electric field is uniform over distances large compared to the cell size so that the interaction with the electrons is limited to the dipolar order. Therefore, the problem of the linear response of the systems studied in this article is similar to that of the polarization of atoms and molecules and the concept of polarizability can be used in our case too. Recall that the polarizability is the ratio of the amplitude of the induced dipole to that of the applied ac electric field. Due to the relatively large cell size compared to that of atoms and to the relatively weak excitation energies, the systems proposed in this article are expected to exhibit large electrical polarizabilities. 

The detailed study of the interaction of a whole array with an electromagnetic wave lies beyond the purpose of the present work. Here, our aim is to obtain the linear response of an isolated center, occupied by a single electron. We use a quasi-static approximation to describe the interaction of this electron with the oscillating electric field. Within this approximation, the electric field obeys the laws of electrostatics although it is derived from a time-dependent electric potential. This results in neglecting any retardation effect, which seems justified as the cell dimensions are far shorter than the radiation wavelength in the THz range. The equivalence between this approximation and the usual semiclassical theory of radiation is discussed in the comment of complement $\text{A}_\text{XIII}$, Sec. 1c of Ref.\ \onlinecite{tan}.
%
%
\subsection{Effect of the central metal wire on the interaction with the electric field}
\label{mwire}
In the present article, we do not take the electrical screening due to the wires at the square vertices into account. This is consistent with studying an electron bound to an isolated wire. However, we include the effect of the central metallic wire on the electron response function. Indeed, the electron in the state bound around the central wire induces a charge distribution at the surface of this latter which brings a non-negligible contribution to the interaction with the oscillating electric field. The easiest way to take this contribution into account is to use the image-charge method. As a first approximation, we consider that the electron charge is distributed uniformly along the $Oz$ direction. Then the problem reduces to finding the image of a charge wire parallel to a conducting cylinder or, equivalently, the 2D image of a point charge with respect to a circle at the ground potential. The solution is well known. See, e.g., Ref.\ \onlinecite{lac}, Sec. 3.5. The image has the same absolute value than the source, but the opposite sign. It is located on the same circle radius, inside the circle, and at the distance $R^2 / r$ from the center. Recall that we use the notation $R$ for the wire radius and $r = \left( x^2 + y^2 \right)^{1/2}$ for the distance to the wire axis. As pointed out above, due to the uniformity of the applied electric field over distances far larger than the cell dimension, the interaction reduces to the dipole contribution. We only consider the case of an electric field in the plane of the semiconductor film. Then, the dipole component of interest is that parallel to this plane so that the electron dipole operator responsible for the interaction with the field is $\hat{\mathbf d}_e = -e \, \mathbf r$ with $\mathbf r = (x,y)$ being the in-plane position of the electron, while that associated with the image charge, according to the properties of this latter described above, is $\hat{\mathbf d}_i = e \left( R^2 / r^2 \right) \mathbf r$. Therefore, the total electric-dipole operator is
\begin{equation}
\hat{\mathbf d} = - e \left( 1 - \frac {R^2} {r^2} \right) \mathbf r
\label{dp1}
\end {equation}
and the Hamiltonian of interaction between the charge in the cell and the external oscillating field in the quasi-static approximation is
\begin{equation}
H_i = e \eta(\mathbf r) \delta \mathcal{E}(t).
\label{dp2}
\end {equation}
In this equation, Eq.\ (\ref{dp2}), $\delta \mathcal{E}(t)$ denotes the externally applied electric field, which is taken oriented along the $Ox$ axis and is written as
\begin{equation}
\delta \mathcal{E}(t) = \delta \mathcal{E}_0 \cos \omega t,
\label{dp3}
\end {equation}
while the notation $\eta(\mathbf r)$ is used for
\begin{equation}
\eta(\mathbf r) = \left( 1 - \frac {R^2} {r^2} \right) x.
\label{dp4}
\end {equation}
Recall that the quasi-static electric field acting upon the cell, $\delta \mathcal{E}(t)$, is either an external uniform field applied parallel to the film plane or that of an electromagnetic wave with its polarization vector in this plane.

Notice that the presence of the metal wire leads to a decrease in the total electric dipole of the cell as well as in the strength of interaction. On the contrary, metal nanocomponents of different shapes have been used to increase the efficiency of light emission in semiconductor nanostructure devices. For more detail, see, e.g., Ref.\ \onlinecite{dong,li,jain} and the references therein. Also, the enhancement of Raman scattering has been known for some time and is frequently used in spectroscopy. See the review by Moskovits.\cite{mosk}  However, in all these cases, the photon energy is in the eV range, close to the resonance with the surface plasmons of the metal. The frequencies implied in the systems under study in the present article are about 100 times lower, far from the plasmon frequencies in the metal wires. This, together with the difference in the system geometries, explains why, in the present case, the presence of the metal has an opposite effect on the interaction with the electromagnetic radiation.
%
%
\subsection{Modified transition matrix elements, oscillator strengths, and sum rule}
\label{matel}
The results of the previous section show that, due to the presence of the central metallic wire, the electric-dipole operator is no longer $- e x$ as in an isolated atom, but $-e \eta(\mathbf r)$. Therefore, the probability of radiative transition between the states $i$ and $j$
%
%
\begin{table}[t]
\renewcommand{\baselinestretch}{1}
\caption{Oscillator-strength sum rules for the radiative transitions involving the ground state, in the case of the two models described in the present article.}
\normalsize
\begin{ruledtabular}
\begin{tabular}{cc}
Model No.&$S_0 = \sum_j f_{0,j}$\\
\noalign{\vskip 1mm}
\hline
\noalign{\vskip 1mm}
1 ($R = \SI{4}{nm}$, $a = \SI{30}{nm}$)&1.219\\
2 ($R = \SI{40}{nm}$, $a = \SI{200}{nm}$)&1.718\\
\end{tabular}
\end{ruledtabular}
\label{srule}
\end{table}
%
%
is now governed by the square of the matrix element
\begin{equation}
t_{i,j} = \iint u_j(\mathbf r) \eta(\mathbf r) u_i(\mathbf r) dx dy.
\label{mt1}
\end {equation}

The concept of oscillator strength is important in atomic and molecular physics. For the meaning of this concept, see, e.g., Chap. 4, Sec. 9\textsuperscript{4} of Ref.\ \onlinecite{con} and Chap XIII, Complement A\textsubscript{XIII}, Sec. 2C of Ref.\ \onlinecite{tan}. It allows the comparison between different radiative transitions of individual electrons. This justifies our interest in computing the oscillator strengths of the systems studied here. Obviously, the transition matrix elements to be used are those given by Eq.\ (\ref{mt1}). This leads to the following modified form of the oscillator strengths
\begin{equation}
f_{i,j} = \frac {2 m_b \left(\epsilon_j - \epsilon_i \right)} {\hbar^2} \lvert t_{i,j} \rvert ^2, \qquad \epsilon_j > \epsilon_i,
\label{mt2}
\end {equation}
with $t_{i,j}$ defined in Eq.\ (\ref{mt1}). These modified oscillator strengths are dimensionless as the oscillator strengths used in atomic physics. They obey a sum rule slightly different from the Thomas-Reiche-Kuhn one. Indeed, in Appendix~\ref{sumrule}, we show that the sum $S_i = \sum_j f_{i,j}$ simply reduces to
\begin{equation}
S_i = \iint u_i^2(\mathbf r) \lvert \nabla \eta(\mathbf r) \rvert^2 \, dx dy.
\label{mt3}
\end {equation}
%

%
\begin{figure}[t]
\includegraphics{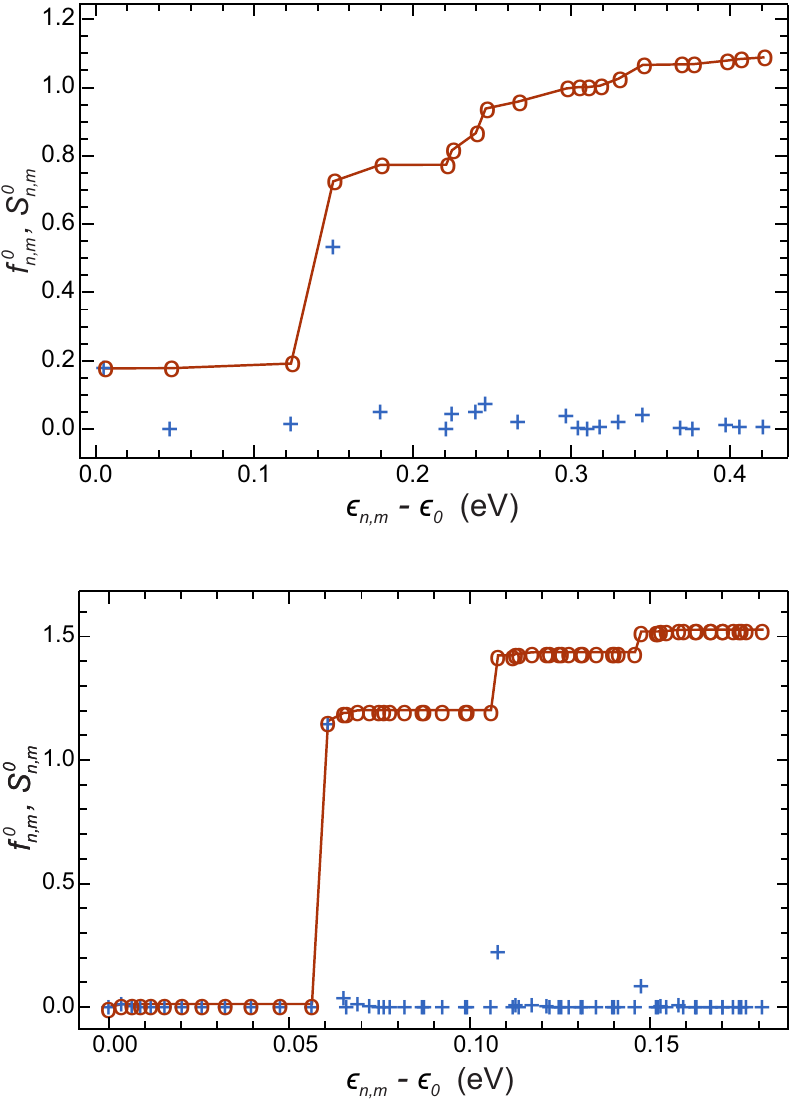}
\renewcommand{\baselinestretch}{1}
\caption{Blue crosses: Oscillator strengths $f^0_{n,m}$; brown circles and full line: Cumulative sums $S^0_{n,m}$ of these oscillator strengths; both for radiative transitions from the ground state to states with $n$ and $m$ as quantum numbers and versus the excitation energy, $\epsilon_{n,m} - \epsilon_0$. Top: model 1, with $R = \SI{4}{nm}$, $a = \SI{30}{nm}$; bottom: model 2, with $R = \SI{40}{nm}$, $a = \SI{200}{nm}$. Notice that just a few oscillator strengths are dominant in the low-energy range considered in the graphs. Compare with the values of the sum rule given in Table~\ref{srule}.}
\label{oscstr}
\end{figure}
\normalsize
%
Most often in atomic physics, the oscillator strengths referred to are those related to the ground state. We comply with this habit and restrict ourselves to the case of transitions involving the ground state. We denote by $f^0_{n,m}$ the oscillator strengths of the transitions from the ground state to the states with quantum numbers $n$ and $m$, $S^0_{n,m}$ their cumulative sum up to $n$ and $m$, and $S_0$ the value given by the sum rule as the limit of $S^0_{n,m}$ for both $n \rightarrow \infty$ and $m \rightarrow \infty$. Using the eigenfunctions obtained in Sec.~\ref{res} and Eqs.\ (\ref{mt1}) and (\ref{mt2}), we compute these oscillator strengths for the two models chosen before. We also apply Eq.\ (\ref{mt3}) to their ground-state wave functions to obtain the values of $S_0$ predicted by our modified sum rule. For both models, the applied bias is \SI{1}{V}. The results of the sum rule are reported in Table~\ref{srule}, while Fig.~\ref{oscstr} shows, in both models, the oscillator strengths for transitions from the ground state as well as their cumulative sum, versus the excitation energy. The ground state being totally symmetric with respect to reflections on both the $Ox$ and $Oy$ axes, the electric-dipole transitions to the states antisymmetric to $Ox$ and symmetric to $Oy$ are the only allowed ones and we do not represent the other transitions. Figure~\ref{oscstr} shows that a few transitions are largely dominant. The dominant transitions occur from the ground state to states with $m = 1$ in the successive shells $n = 2, 3, \ldots$. These states have a node along the $Oy$ axis. This geometry explains why they produce a large electric dipole along the $Ox$ axis and, therefore, why the corresponding oscillator strengths are dominant. The optical spectra of the systems studied here should show well-defined peaks at the frequencies of these transitions. We confirm this conclusion in the next section.
%
%
\subsection{Polarizability and absorption spectrum}
\label{spectr}
We now turn to computing the polarizability and the absorption spectrum of an isolated cell with an electron in the ground state. These are important properties for the purpose of possible electromagnetic applications. Again, we assume that the intensity of the radiation field is weak and its wavelength large compared to the cell dimensions so that we can restrict the computations to the linear dipolar term in the interaction between the electron and the electromagnetic field. Therefore, the quantum-mechanical average, or expectation value, of the electron dipole momentum, $d(t)$, appears as proportional to the amplitude of the electric field, $\delta \mathcal{E}_0$. This dipole momentum has two components, an in-phase component, $d_1(t)$, and a quadrature component, $d_2(t)$, such that
\begin{subequations}
\label{pl1}
\begin{eqnarray}
d_1(t) = \alpha_1(\omega) \, \delta \mathcal{E}_0 \cos(\omega t),
\label{pl1a}
\\
d_2(t) = \alpha_2(\omega) \, \delta \mathcal{E}_0 \sin(\omega t).
\label{pl1b}
\end{eqnarray}
\end{subequations}

By analogy with the case of atoms and molecules, we call polarizability the property which has $\alpha_1(\omega)$ and $\alpha_2(\omega)$ as components. A complex electric field $\delta \mathcal{E}_0 \, e^{i \omega t}$ could be used instead of the real one of Eq.\ (\ref{dp3}), leading to a complex expression of the polarizability with $\alpha_1(\omega)$ and $- \alpha_2(\omega)$ as real and imaginary parts, respectively. It is well known that the quadrature, or imaginary, component is in direct connection with the power absorbed from the electromagnetic radiation. Indeed, if $d \left[ d_1(t) + d_2(t) \right]$ is the change of the system dipole momentum in the time interval $dt$, then the power absorbed by the electron at the time $t$ is
\begin{eqnarray}
\mathcal{P}(\omega,t) = \delta \mathcal{E}^2_0 \, \omega \left[ \alpha_1(\omega) \sin(\omega t) \cos(\omega t) \right.
\nonumber
\\
\left.  - \alpha_2(\omega) \cos^2(\omega t) \right].
\label{pl2}
\end{eqnarray}

The quantity of interest here is the power averaged over a period, which we call $\mathcal{P}(\omega)$. The first term in brackets in Eq.\ (\ref{pl2}) has a time average equal to zero. Therefore, the net absorbed power is due to the existence of a quadrature term in the polarizability,
\begin{equation}
\mathcal{P}(\omega) = - \frac {1} {2} \, \alpha_2(\omega) \, \omega \, \delta \mathcal{E}^2_0.
\label{pl3}
\end {equation}
Similarly in solid-state physics, the imaginary part of the dielectric constant gives the attenuation of an electromagnetic wave propagating in an absorbing medium. For a detailed description, see, e.g., Ref.\ \onlinecite{kling}. 

To compute the polarizability, we use the semiclassical theory of radiation and treat its interaction with the electron in the framework of the quasi-static approximation. The time-dependent perturbation theory suits our purpose of obtaining the linear response of the system studied here. See, e.g., Chap. XIII and its complements in Ref.\ \onlinecite{tan} for a description of the perturbation theory applied to the interaction between electrons and the electromagnetic radiation. Recall that we take the ac electric field acting upon the electron oriented along the $Ox$ axis. The electric-dipole operator, including the contribution of the image charge in the metal wire, is that of Eq.\ (\ref{dp1}). By reason of symmetry, its quantum-mechanical average is also aligned with the $Ox$ axis so that we only need its $x$ component which is $-e \eta(\mathbf r)$.  The unperturbed Hamiltonian $H_0$, defined in Eq.\ (\ref{md1a}), is the one used in the discussion of the eigenstates in Sec.~\ref{res} and its eigenvalues are denoted by $\epsilon_i$. Here, for the sake of easiness, we use a single index $i$ to represent the quantum numbers $n$ and $m$.

The perturbation theory gives for the ground-state wave function restricted to the linear order
\begin{equation}
\phi_0(\mathbf r, t) = u_0(\mathbf r) + W(t) u_0(\mathbf r)
\label{pl4}
\end {equation}
with
\begin{eqnarray}
W(t) = - \frac {e \delta \mathcal{E}_0} {2} \left(\frac {e^{i (H_0 - \epsilon_0 - \hbar \omega) t / \hbar}} {H_0 - \epsilon_0 - \hbar \omega +i \hbar \lambda} \right.
\nonumber
\\
\left. + \frac {e^{i (H_0 - \epsilon_0 + \hbar \omega) t / \hbar }} {H_0 - \epsilon_0 + \hbar \omega +i \hbar \lambda} \right) \eta (\mathbf r).
\label{pl5}
\end{eqnarray}
We have added imaginary terms to the denominators to describe the effects of the finite lifetime of the states reached in the transitions. The parameter $\lambda$ is the inverse of the relaxation time $\tau$ of the electrons in these states, $\lambda = 1/ \tau$. The causes of decay are numerous, for example radiative decay, phonon emissions, collisions with defects, Auger effects, etc. Their importance in the present case is not known. As there are no experimental results to compare with, the chosen value of the relaxation time is not crucial. We suggest to use $\tau = \SI{0.1}{ps}$, which is in the range of the collision times deduced from bulk-mobility measurements. See the discussion in Sec.~\ref{mot}.

Using Eqs.\ (\ref{pl4}) and (\ref{pl5}), we obtain the expectation value of the system electric dipole computed at the linear order in $\delta \mathcal{E}_0$,
\begin{eqnarray}
d(t) &=& - \frac {e^2} {2} \delta \mathcal{E}_0 \iint u_0(\mathbf r) \eta(\mathbf r) \left( \frac {e^{-i \omega t}} {H_0 - \epsilon_0 - \hbar \omega +i \hbar \lambda} \right.
\nonumber
\\
&& + \left. \frac {e^{i \omega t}} {H_0 - \epsilon_0 + \hbar \omega +i \hbar \lambda} \right) \eta(\mathbf r) u_0(\mathbf r) \, dx dy + \text{cc}.
\nonumber
\\
\label{li4}
\end{eqnarray}
Call
\begin{equation}
D(\mathbf r, \omega) = - \frac {1} {H_0 - \epsilon_0 - \hbar \omega +i \hbar \lambda} \, \eta(\mathbf r) u_0(\mathbf r)
\label{li5}
\end {equation}
and $D_1(\mathbf r, \omega) = \text{Re} D(\mathbf r, \omega)$, $D_2(\mathbf r, \omega) = \text{Im} D(\mathbf r, \omega)$. Then, the polarizability components can be written as
\begin{subequations}
\label{li6}
\begin{eqnarray}
\alpha_1(\omega) &=& - e^2 \iint \eta(\mathbf r) u_0(\mathbf r) \left[ D_1(\mathbf r, \omega) \right.
\nonumber
\\
&& \left. + D_1(\mathbf r, -\omega) \right] \, dx dy,
\label{li6a}
\\
\alpha_2(\omega) &=& - e^2 \iint \eta(\mathbf r) u_0(\mathbf r) \left[ D_2(\mathbf r, \omega) \right.
\nonumber
\\
&& \left. - D_2(\mathbf r, -\omega) \right] \, dx dy.
\label{li6b}
\end{eqnarray}
\end{subequations}
In case of no damping, i.e., if $\lambda = 0$, the imaginary part of $D(\mathbf r, \omega)$ is zero as well as the average absorbed power, as expected.

Introducing a complete set $u_j(\mathbf r)$ of eigenfunctions of $H_0$, we can write the polarizability components as
\begin{subequations}
\label{li6-1}
\begin{eqnarray}
\alpha_1(\omega) &=& e^2 \sum_j \eta_{0,j}^2 \text{Re} \left[ \frac {1} {\epsilon_j - \epsilon_0 - \hbar \omega +i \hbar \lambda} \right.
\nonumber
\\
&& \left. + \frac {1} {\epsilon_j - \epsilon_0 + \hbar \omega +i \hbar \lambda} \right],
\label{li6-1a}
\\
\alpha_2(\omega) &=& e^2 \sum_j \eta_{0,j}^2 \text{Im} \left[ \frac {1} {\epsilon_j - \epsilon_0 - \hbar \omega +i \hbar \lambda} \right.
\nonumber
\\
&& \left. - \frac {1} {\epsilon_j - \epsilon_0 + \hbar \omega +i \hbar \lambda} \right],
\label{li6-1b}
\end{eqnarray}
\end{subequations}
in which equations $\eta_{0,j}$ denotes the matrix element
\begin{equation}
\eta_{0,j} = \iint \eta(\mathbf r) u_0(\mathbf r) u_j(\mathbf r) dx dy.
\label{si4}
\end {equation}
%
%
\begin{figure}[b]
\includegraphics{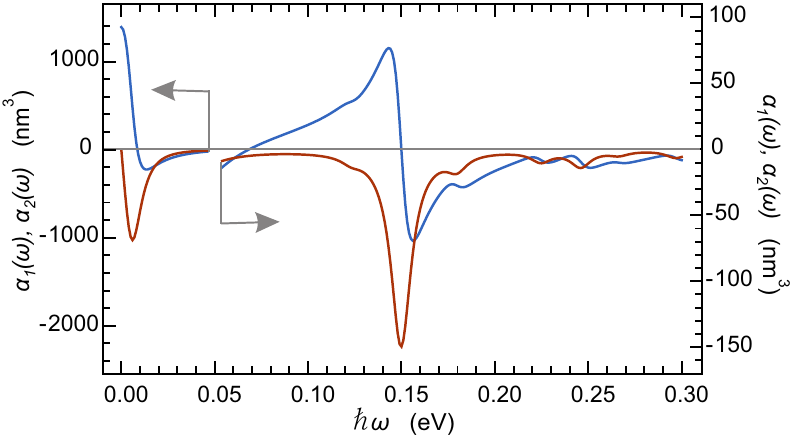}
\renewcommand{\baselinestretch}{1}
\caption{Blue line: in-phase component, $\alpha_1(\omega)$, and red line: quadrature component, $\alpha_2(\omega)$, of polarizability in \si{nm^3} for model 1, versus photon energy, $\hbar \omega$. Notice the two scales, on left $y$ axis for $\hbar \omega < \SI{0.05}{eV}$ and on right $y$ axis for $\hbar \omega > \SI{0.05}{eV}$.}
\label{twpol}
\end{figure}
\normalsize
%
This allows to compute the polarizability term by term. Recall that the transitions from the ground state to the states which are not antisymmetric to $Ox$ and symmetric to $Oy$ are symmetry forbidden. However, the complete polarizability can be computed at once in the following way. From Eq.\ (\ref{li5}) it can be deduced that $D(\mathbf r,\omega)$ obeys the following partial differential equation
\begin{equation}
(H_0 - \epsilon_0 - \hbar \omega +i \hbar \lambda) D(\mathbf r,\omega) = - \eta(\mathbf r) u_0(\mathbf r).
\label{li10}
\end {equation}
Separating the real and imaginary parts of $D(\mathbf r,\omega)$ in Eq.\ (\ref{li10}) leads to a system of two real partial differential equations, namely
\begin{subequations}
\label{li11}
\begin{eqnarray}
(H_0 - \epsilon_0 - \hbar \omega) D_1(\mathbf r,\omega) - \hbar \lambda D_2(\mathbf r,\omega) &=& - \eta(\mathbf r) u_0(\mathbf r),
\nonumber
\\
\label{li11a}
\\
(H_0 - \epsilon_0 - \hbar \omega) D_2(\mathbf r,\omega) + \hbar \lambda D_1(\mathbf r,\omega) &=& 0.
\label{li11b}
\end{eqnarray}
\end{subequations}
This system of equations is solved in COMSOL with boundary conditions appropriate to the symmetry of $\eta(\mathbf r) u_0(\mathbf r)$ and the results are used in Eqs.\ (\ref{li6-1}) to obtain the polarizability components. The spectrum of absorbed power, $\mathcal{P}(\omega)$, is easily deduced from the values obtained for $\alpha_2(\omega)$ by means of Eq.\ (\ref{pl3}).

%
\begin{figure}[t]
\includegraphics{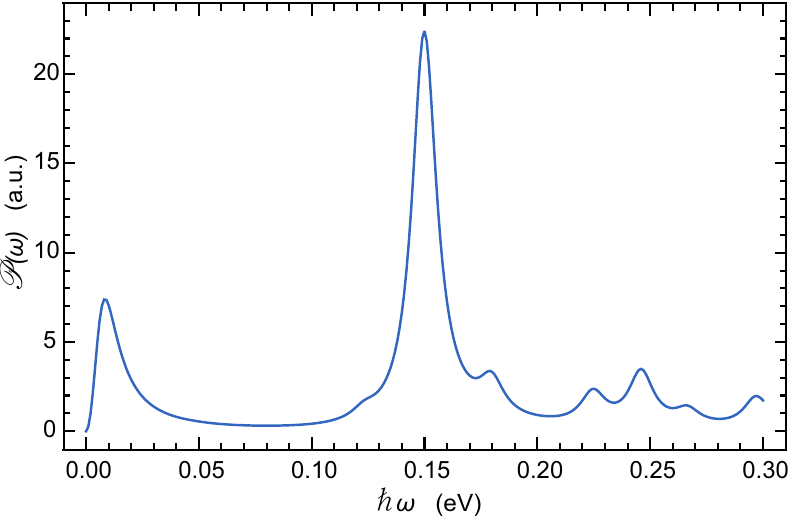}
\renewcommand{\baselinestretch}{1}
\caption{Absorbed power in arbitrary units, $\mathcal{P}(\omega)$, in the case of model 1, versus photon energy, $\hbar \omega$, in \si{eV}.}
\label{twabspw}
\end{figure}
\normalsize
%
%
\begin{figure}[b]
\includegraphics{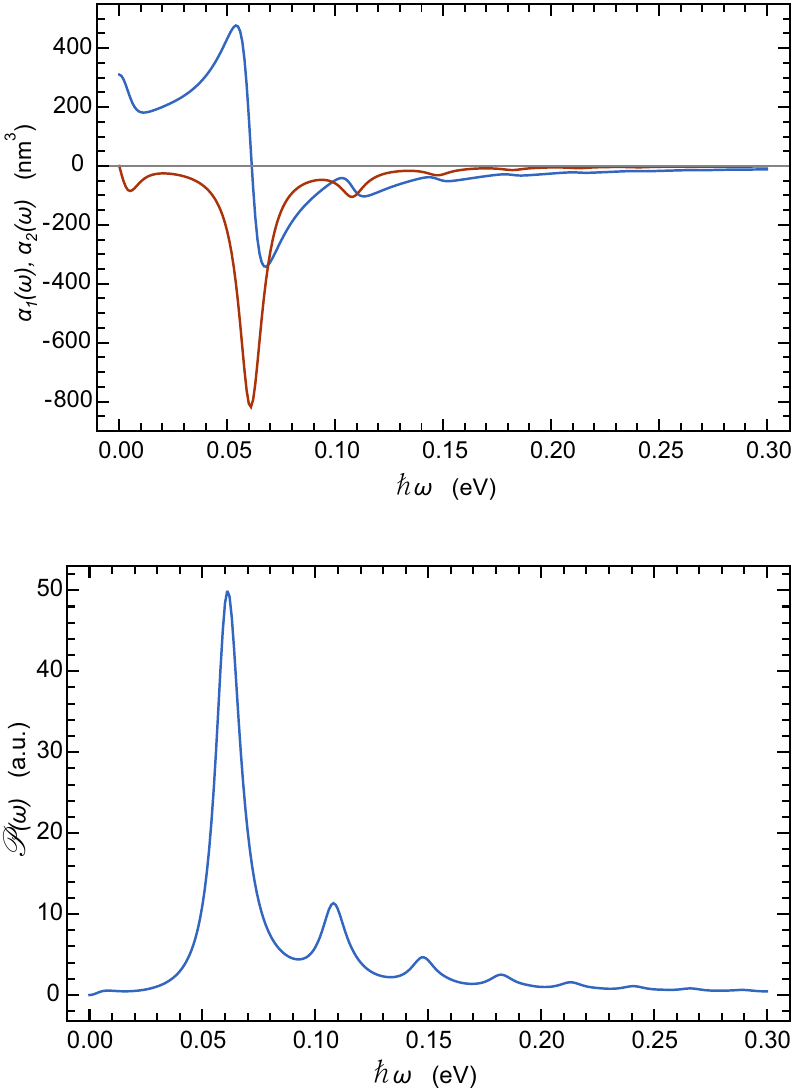}
\renewcommand{\baselinestretch}{1}
\caption{Model 2. Top: Blue line: in-phase component, $\alpha_1(\omega)$, and red line: quadrature component, $\alpha_2(\omega)$, of polarizability in \si{nm^3}, versus photon energy, $\hbar \omega$. Bottom: Absorbed power in arbitrary units, $\mathcal{P}(\omega)$, also versus photon energy, $\hbar \omega$, in \si{eV}.}
\label{lwpolpow}
\end{figure}
\normalsize
%
First, consider model 1 with $R = \SI{4}{nm}$, $a = \SI{30}{nm}$, and the bias voltage $V_B = \SI{1}{V}$. Figure~\ref{twpol} shows the in-phase and quadrature components of the polarizability obtained as explained above, versus the photon energy of the electromagnetic radiation. Recall that a frequency of $\SI{1}{THz}$ corresponds to about $\SI{4}{meV}$. The contribution due to the transition to the $n = 1, m = 1$ state, in the same shell as the initial state, is largely dominant at low photon energy, reaching values above $\SI{1000}{nm^3}$. This leads us to use two different scales in the graph of Fig.~\ref{twpol}, a first one for photon energies below $\SI{50}{meV}$, a second one for photon energies above $\SI{50}{meV}$. The resonance next in frequency appears at about $\SI{150}{meV}$, i.e., in the mid-infrared region of the electromagnetic radiation. Close to this frequency, the polarizability reaches a large value of nearly $\SI{100}{nm^3}$. This resonance is due to the transition to the $n = 2, m = 1$ state. This large value of the polarizability is in agreement with that of the corresponding oscillator strength, $f_{2,1}^0 = 0.533$, shown in the top graph of Fig.~\ref{oscstr}.

%
\begin{figure}[b]
\includegraphics{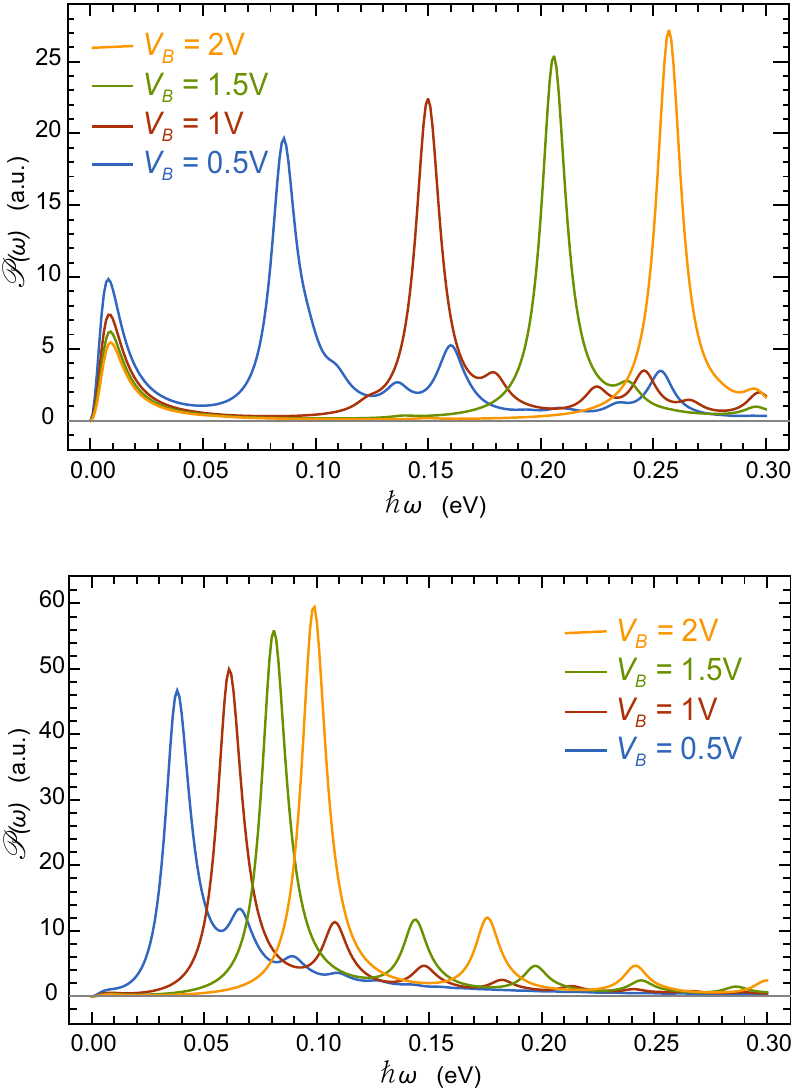}
\renewcommand{\baselinestretch}{1}
\caption{Absorbed power in arbitrary units, $\mathcal{P}(\omega)$ versus photon energy, $\hbar \omega$, in \si{eV} for different bias voltages, $V_B = \SI{0.5}{V}$, $V_B = \SI{1}{V}$, $V_B = \SI{1.5}{V}$, and $V_B = \SI{2}{V}$. Top figure: case of model 1; bottom figure: case of model 2;}
\label{powabsvsV}
\end{figure}
\normalsize
%
Figure~\ref{twabspw} shows the product $- \alpha_2(\omega) \hbar \omega$ which, according to Eq.\ (\ref{pl3}), is a measure, in arbitrary units, of the power absorbed from the electromagnetic field in the cell under consideration. The contribution from the transition to the $n = 1, m = 1$ state is strongly reduced in the power spectrum, due to the presence of the factor $\omega$ in the definition of $\mathcal{P}$. The dominant transition now is that to the $n = 2, m = 1$ state at about $\SI{150}{meV}$.

The upper graph in Fig.~\ref{lwpolpow} shows both components of the polarizability in the case of model 2 ($R = \SI{40}{nm}$, $a = \SI{200}{nm}$) while the lower graph shows $- \alpha_2(\omega) \hbar \omega$, i.e., the absorbed power in arbitrary units in the same model. The resonance occurring at about \SI{52}{meV}, due to the transition to the state $n = 2, m = 1$, gives a contribution largely dominant in the polarizability as well as in the absorbed power.

In both models, the resonance in the polarizability and the main peak in the absorbed-power spectrum due to the transition to the $n = 2, m = 1$ state are relatively sharp and well defined. This makes the systems discussed in the present article good candidates for the development of new ir detectors, notwithstanding the possible difficulties in setting them up practically.

Figure~\ref{powabsvsV} shows the product $- \alpha_2(\omega) \hbar \omega$, i.e., the absorbed power $\mathcal{P}(\omega)$ vs photon energy $\hbar \omega$ for four different bias voltages from $\SI{0.5}{V}$ to $\SI{4}{V}$. The upper and lower graphs are related to models 1 and 2, respectively. The figure shows that if photodetectors based on the present systems could be built, they could be easily made tunable in a range from $\SI{10}{THz}$ to $\SI{100}{THz}$ approximately, in the case of model 1 and from $\SI{5}{THz}$ to $\SI{30}{THz}$ in that of model 2.
%
%
%
\section{Conclusions}
\label{concl}
In the systems studied in the present article electrons occupy quantum states around metal nanowires which are embedded in a semiconductor film and oriented normal to the plane surface of this latter. These systems have the interesting property that the voltage applied to the wires modify the electron excitation energies leading to tunable spectra for the interaction with infrared electromagnetic waves. As shown in this article for the absorption spectrum, the frequency range over which the position of the absorption peaks can be tuned is large. This makes these systems good candidates for the development of new detectors or emitters of THz radiation, notwithstanding the difficulty of practical realization. A difficulty could come from the perturbation brought to the electromagnetic wave by the presence of metallic wires. The article shows how to include the effect of the central wires in the calculation of the oscillator strengths and of the absorption spectrum. However, that of the wires at the cell vertices are not taken into account. Replacing the metal wires by undoped or slightly doped semiconductor cylinders would solve this problem. The theory of such systems is part of our program of future research.
%
%
%
\appendix
%
%
%
%
\section{Sum rule}
\label{sumrule}
The purpose of this appendix is to obtain the equivalent of the Thomas-Reiche-Kuhn sum rule if the image dipole induced in the central metal wires is taken into account. The expression of the oscillator strength for the transition from the state $i$ to the state $j$, including the contribution of the image dipole, is given by Eq.\ (\ref{mt2}) in Sec.~\ref{matel}. It is
\begin{equation}
f_{i,j} = \frac {2 m_b \left(\epsilon_j - \epsilon_i \right)} {\hbar^2} \lvert t_{i,j} \rvert ^2,
\label{sr1}
\end {equation}
in which equation $t_{i,j}$ denotes the matrix element
\begin{equation}
t_{i,j} = \iint u_j(\mathbf r) \eta(\mathbf r) u_i(\mathbf r) dx dy.
\label{sr2}
\end {equation}
Obviously, the oscillator strengths $f_{i,j}$ have no diagonal term. Recall that $-e \eta(\mathbf r)$ is the electron electric-dipole operator including the contribution of the image charge, while $u_i(\mathbf r)$ and $\epsilon_i$ denote the eigenfunctions and the eigenvalues of the 2D Hamiltonian $H_0$ defined in Eq.\ (\ref{md1}) in Sec.~\ref{form}, respectively. These eigenfunctions are real and normalized over the cell surface.

Consider the sum $S_i = \sum_j f_{i,j}$. Using Eq.\ (\ref{sr1}) and the fact that the eigenstates constitute a complete set, we write this sum as 
\begin{equation}
S_i = \frac {2 m_b} {\hbar^2} \iint u_i(\mathbf r) \eta(\mathbf r) \left[ H_0, \eta(\mathbf r) \right] u_i(\mathbf r) \, dxdy.
\label{sr3}
\end {equation}
But $[H_0, \eta(\mathbf r)] = (1 / 2m_b) \left \{ \mathbf p \cdot [ \mathbf p, \eta(\mathbf r)] + [ \mathbf p, \eta(\mathbf r)] \cdot \mathbf p \right \}$ and $[\mathbf p , \eta(\mathbf r)] = -i \hbar \bm{\nabla} \eta(\mathbf r)$, so that, successively,
\begin{eqnarray}
S_i = - \frac {i} {\hbar} \iint u_i(\mathbf r) \eta(\mathbf r) \left \{ \mathbf p \cdot \bm{\nabla} \eta(\mathbf r) \right.
\nonumber
\\
 + \left. \bm{\nabla} \eta(\mathbf r) \cdot \mathbf p \right \} u_i(\mathbf r) \, dxdy,
\label{sr3b}
\end{eqnarray}
and
\begin{eqnarray}
S_i &=& \iint u_i(\mathbf r)  \lvert \nabla \eta(\mathbf r) \rvert ^2 u_i(\mathbf r) \, dxdy
\nonumber
\\
&& - \frac {i} {\hbar} \iint u_i(\mathbf r) \mathbf p \cdot \eta(\mathbf r) \bm{\nabla} \eta(\mathbf r) u_i(\mathbf r) \, dxdy
\nonumber
\\
&& - \frac {i} {\hbar} \iint u_i(\mathbf r) \eta(\mathbf r) \bm{\nabla} \eta(\mathbf r) \cdot \mathbf p u_i(\mathbf r) \, dxdy.
\label{sr4}
\end{eqnarray}

In the third integral of Eq.\ (\ref{sr4}) $\mathbf p u_i(\mathbf r)$ leads to $-i \hbar \bm{\nabla} u_i(\mathbf r)$, while, in the second one, $u_i(\mathbf r) \mathbf p$ gives $i \hbar \bm{\nabla} u_i(\mathbf r)$ so that these two integrals cancel each other out and we are left with the result announced in Eq.\ (\ref{mt3}), Sec.~\ref{matel},
\begin{equation}
S_i = \iint u_i^2(\mathbf r) \lvert \nabla \eta(\mathbf r) \rvert^2 \, dx dy
\label{sr5}
\end {equation}
which constitutes a modified Thomas-Reiche-Kuhn sum rule. In fact, the oscillator strengths with $\epsilon_j < \epsilon_i$ have no actual meaning. Because, in this article, we consider transitions from the ground state only, this poses no problem here. In other cases, the contributions of the terms with $\epsilon_j < \epsilon_i$ should be subtracted from the result of Eq.\ (\ref{sr5}).

%
\bibliography{warray}
\end{document}